\documentclass[
 aps,
 prx,
 amsmath,amssymb,
 superscriptaddress,
 preprint,
 ]{revtex4-2}

\usepackage{graphicx}
\usepackage{dcolumn}
\usepackage{bm}
\usepackage{xcolor}
\usepackage{float}
\usepackage{soul}

\begin{document}

\title{Intrinsic Limits of Charge Carrier Mobilities in Layered Halide Perovskites}

\author{Bruno Cucco}%
\affiliation{Univ Rennes, ENSCR, CNRS, ISCR (Institut des Sciences Chimiques de Rennes)– UMR 6226, F-35000 Rennes, France.}%
\affiliation{Oden Institute for Computational Engineering and Sciences, The University of Texas at Austin, Austin, Texas 78712, USA
and Department of Physics, The University of Texas at Austin, Austin, Texas 78712, USA.}%
\author{Joshua Leveillee}%
\affiliation{Oden Institute for Computational Engineering and Sciences, The University of Texas at Austin, Austin, Texas 78712, USA
and Department of Physics, The University of Texas at Austin, Austin, Texas 78712, USA.}%
\author{Viet-Anh Ha}%
\affiliation{Oden Institute for Computational Engineering and Sciences, The University of Texas at Austin, Austin, Texas 78712, USA
and Department of Physics, The University of Texas at Austin, Austin, Texas 78712, USA.}%
\author{Jacky Even}%
\affiliation{Univ Rennes, INSA Rennes, CNRS, Institut FOTON - UMR 6082, F-35000 Rennes, France.}%
\author{Mika\"el Kepenekian}%
\affiliation{Univ Rennes, ENSCR, CNRS, ISCR (Institut des Sciences Chimiques de Rennes)– UMR 6226, F-35000 Rennes, France.}%
\email{mikael.kepenekian@univ-rennes.fr}
\author{Feliciano Giustino}%
\affiliation{Oden Institute for Computational Engineering and Sciences, The University of Texas at Austin, Austin, Texas 78712, USA
and Department of Physics, The University of Texas at Austin, Austin, Texas 78712, USA.}%
\author{George Volonakis}%
\email{yorgos.volonakis@univ-rennes.fr}
\affiliation{Univ Rennes, ENSCR, CNRS, ISCR (Institut des Sciences Chimiques de Rennes)– UMR 6226, F-35000 Rennes, France.}%

\begin{abstract}
Layered halide perovskites have emerged as potential alternatives to three-dimensional
halide perovskites due to their improved stability and larger material phase space,
allowing fine-tuning of structural, electronic, and optical properties.
However, their charge carrier mobilities are significantly smaller than that of
three-dimensional halide perovskites, which has a considerable impact on their
application in optoelectronic devices.
Here, we employ state-of-the-art ab initio approaches to unveil the electron-phonon
mechanisms responsible for the diminished transport properties of layered halide perovskites.
Starting from a prototypical ABX$_{3}$ halide perovskite, we model the case of $n=1$
and $n=2$ layered structures and compare their electronic and transport properties
to the three-dimensional reference.
The electronic and phononic properties are investigated within density functional theory
(DFT) and density functional perturbation theory (DFPT), while transport properties
are obtained via the ab initio Boltzmann transport equation.
The vibrational modes contributing to charge carrier scattering are investigated and
associated with polar-phonon scattering mechanisms arising from the long-range
Fr{\"o}hlich coupling and deformation potential scattering processes.
Our investigation reveals that the lower mobilities in layered systems primarily originates from the increased electronic density of states at the vicinity of the band edges, while the electron-phonon coupling strength remains similar. Such increase is caused by the dimensionality reduction and
the break in octahedra connectivity along the stacking direction.
Our findings provide a fundamental understanding of the electron-phonon coupling
mechanisms in layered perovskites and highlight the intrinsic limitations of the
charge carrier transport in these materials.
\end{abstract}

\maketitle

\newpage
Layered halide perovskites emerged as a promising class of materials for optoelectronic applications due to their highly tunable physicochemical properties. These materials can be synthesized with different ordering along the stacking axis and number of inorganic layers \textit{n}, which allows tuning of their optoelectronic properties via both dielectric and quantum confinement effects.\cite{Katan2019,Pedesseau2016,Rueda2017} Their electronic properties can also be fine-tuned by engineering the halogen and metal-site composition.\cite{Matheu2022} Different from many conventional lead-based halide perovskites, the layered perovskites can be highly stable in ambient conditions due to the possibility of introducing hydrophobic spacer cations, allowing significant improvements in their stability while preserving their photo-active properties.\cite{Smith2014,Tsai2016,Smith2018,Zheng2019}

As a state-of-the-art application, layered perovskites have been recently incorporated in 2D/3D perovskite heterostructures for solar cells. These architectures are being currently investigated as a pathway to inherit the enhanced stability of layered systems while maintaining the high power conversion efficiencies of three-dimensional perovskite solar cells.\cite{Byeon2023,Metcalf2023,Sidhik2022} Moreover, layered perovskite-based light-emitting devices were show to allow injecting large current values of several ampere per cm$^{2}$~\cite{Tsai2018} or exhibit high color purity and narrow emission, remarkably for blue phosphors, which is of interest for application in modern display technologies~\cite{Smith2019,Wang2017_2,Kawano2014}. Despite these promising advances, the efficiency of layered perovskite-based optoelectronic devices still severely hindered due to large electronic band gaps, strongly bounded excitons and poor charge carrier mobilities at room temperature~\cite{Mitzi1994,Kober2022}. In particular, charge carrier mobilities are shown to range from 0.36~cm$^{2}$/Vs to 8.7~cm$^{2}$/Vs~\cite{Rueda2017,Motti2023} for layered systems, in contrast to 60.0~cm$^{2}$/Vs for the three-dimensional perovskite CH$_{3}$NH$_{3}$PbI$_{3}$~\cite{Xia2021}. As a result of the poor mobilities of layered systems their application in many optoelectronic technologies is a major limiting factor towards highly efficient optoelectronic devices by comparison to other 2D quantum well-like semiconductors \cite{Blancon2020,Milot2016}. To date, various intrinsic or extrinsic mechanisms have been proposed to explain these reduced mobilities, such as large carrier effective masses~\cite{Rueda2017}, high concentration of trap states that could act as scattering centers~\cite{Buizza2019}, and the presence of strongly bounded charge carriers (large exciton binding energies) that adversely affect charge carrier dissociation~\cite{Kober2022}. Yet, the fundamental atomic-scale mechanisms underlying the measured low mobilities remain elusive because a theoretical investigation of the transport properties, including accurately electron-phonon coupling in metal halide layered materials, is still lacking.

In this work, we unveil the effect of dimensionality crossover from three-dimensional to layered halide perovskites on transport properties.
By employing a combination of density functional theory (DFT) and the state-of-the-art solution of the ab initio Boltzmann Transport Equation (\textit{ai}BTE), we show that the measured decrease in the charge carrier mobilities of layered systems is primarily originating from differences in charge carrier lifetimes. More importantly, by decomposing the different contributions to the charge carrier scattering rates, we demonstrate that the mobility of these systems is intrinsically limited by the signicantly higher density of states in the vicinity of the band edges due to the lower dimensionality. A symmetry analysis of the electron-phonon coupling reveals that polar-phonon scattering arising from the long-range Fr{\"o}hlich coupling and non-polar optical deformation potential scattering are the main sources of charge carrier relaxation in layered perovskite materials.

\subsection*{Structural models}
\begin{figure}[!h]
    \centering
    \includegraphics[width=0.8\linewidth]{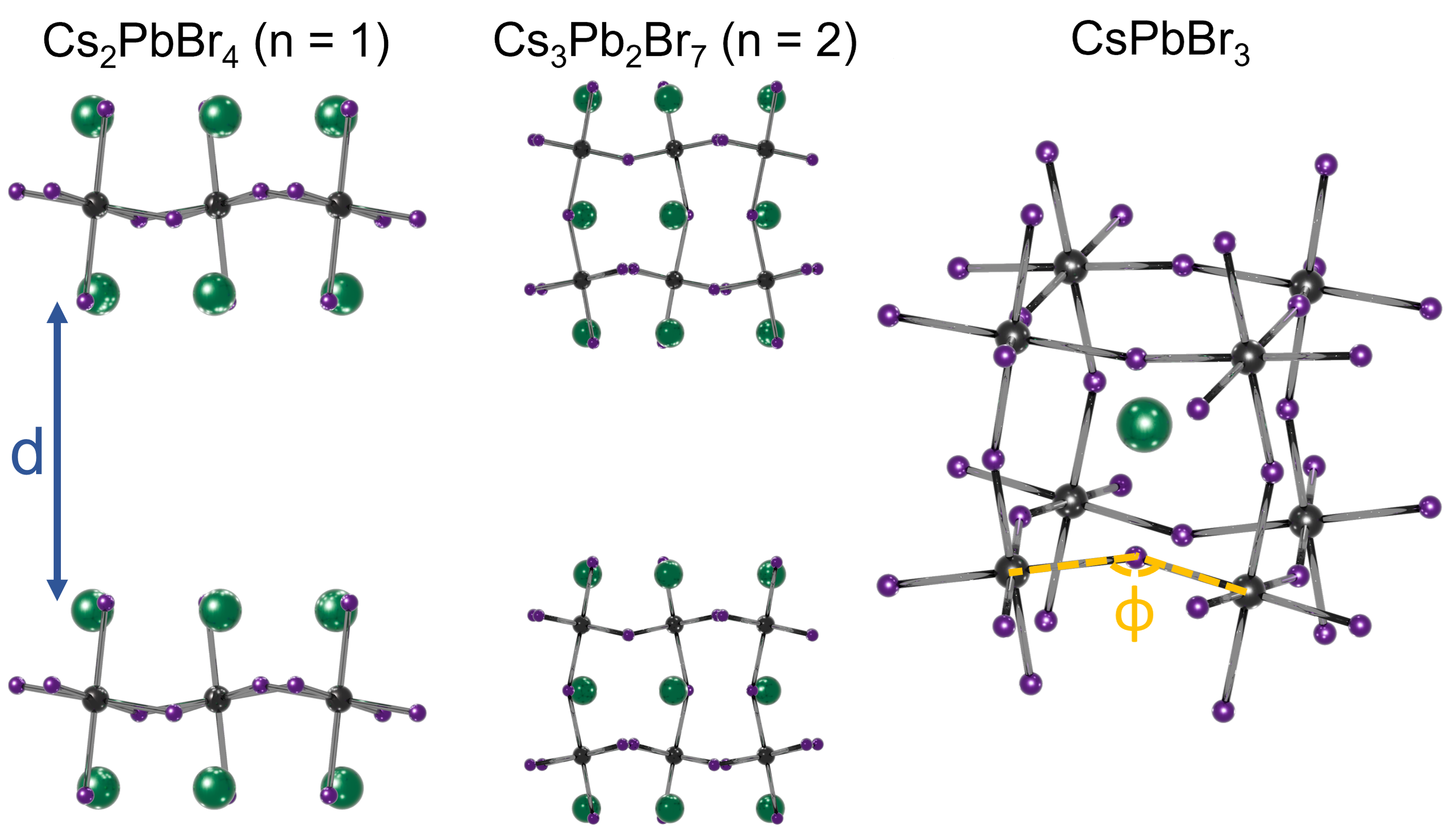}
    \vspace{-0.5cm}
    \caption{%
        Structural models of layered n=1 and n=2, and three-dimensional halide perovskites.
        The interlayer distance \textit{d} is defined as the shortest distance between
        halogens of different layers and is set to 35~\AA~to generate free-standing layers.
        Relaxed ground-state structures show structural parameters (Pb-Br bond length,
        in-plane tilting $\Phi$) in excellent agreement with those of synthesized
        systems.
        }
    \label{fig:structures}
\end{figure}

To explore the structure-property relations in halide perovskites along dimension reduction, we start from the well-characterized three-dimensional compounds CsPbBr$_3$ in its orthorhombic phase~\cite{Mannino2020,Stoumpos2013_2}. We then build free-standing $n=1$ (Cs$_{2}$PbBr$_{4}$) and $n=2$ (Cs$_{3}$Pb$_{2}$Br$_{7}$) model layered perovskites.
This choice of systems with inorganic cations allows us to focus on the effect
of dimension reduction without being influenced by the choice made for large
and small organic cations.
We determine the appropriate ground-state (GS) structures within DFT by displacing the atomic coordinates in the unit cell following the imaginary phonon modes in each model structure. These are obtained via density functional perturbation theory (DFPT) calculations. The set of atomic displacements can be written as:
\begin{equation}
\Delta\chi_{\kappa \alpha p}=\left(\frac{M_{0}}{N_{p}M_{\kappa}}\right)^{1/2}\sum_{\textbf{q}\nu}e^{i\textbf{q}.\textbf{R}_{p}}e_{\kappa\alpha,\nu}(\textbf{q})z_{\textbf{q}\nu},
\end{equation}
where $M_{0}$ is the proton mass, $M_{\kappa}$ is the mass of the $\kappa$th ion, $N_{p}$ is the number of unit cells in the Born–von
Kármán (BvK) supercell, \textbf{q} is the crystal momentum of the lattice vibration, $e_{\kappa\alpha,\nu}(\textbf{q})$ is the
polarization of the acoustic wave corresponding to the wave
vector \textbf{q} and mode $\nu$, and $z_{\textbf{q}\nu}$ are the normal mode coordinates.\cite{Giustino2017}
The GS structures of each material are shown in Figure~\ref{fig:structures}.

The relaxed GS structures of $n=1$, $n=2$ and three-dimensional models
exhibit space-groups P4/mbm (Number 127), P4bm (Number 100) and Pnma (Number 62), respectively.
The average Pb-Br bond lengths are nearly identical in all systems ($3.04$~\AA, $3.04$~\AA{}, and $3.03$~\AA{} for three-dimensional, $n=1$, and $n=2$, respectively),
while the average in-plane tilting (Angle $\Phi$ in Figure~\ref{fig:structures}) is $153.5$\textdegree{}
for CsPbBr$_3$, and $147.5$\textdegree, $155.9$\textdegree~for $n=1$, $n=2$ GS structures, respectively.
The structural properties of the three-dimensional compound are consistent with previously
reported experimental and theoretical data for CsPbBr$_{3}$~\cite{Mannino2020,Stoumpos2013_2}.
In the same manner, the structural parameters obtained for the layered $n=1$ system are in excellent agreement with the experimental data available in the literature. The in-plane tilting angle deviation are of $-2.2\%$, $0.06\%$, $-0.47\%$ and $-0.61\%$ in relation to (PEA)$_{2}$PbBr$_{4}$, (MPenDA)PbBr$_{4}$, (ODA)PbBr$_{4}$ and (BDA)PbBr$_{4}$, respectively.\cite{Smith2017,Ghosh2020} The Pb-Br bond length deviation is of $0.98\%$ in relation to (PEA)$_{2}$PbBr$_{4}$.\cite{Shibuya2009} The in-plane tilting angles are also consistent with the work of Stoumpos \textit{et al.}, where $n=1$ is shown to exhibit larger Pb-Br-Pb tiltings than $n=2$ and three-dimensional MAPbI$_{3}$, which have similar tilting angles.\cite{Stoumpos2016}
This analysis validates our model choice as it confirms that we retain the essential
characteristics of the octahedra network and its evolution going from three-dimensional to
layered systems.

\subsection*{Phonons and Electrons}
\begin{figure*}[ht]
    \centering
    \includegraphics[width=\linewidth]{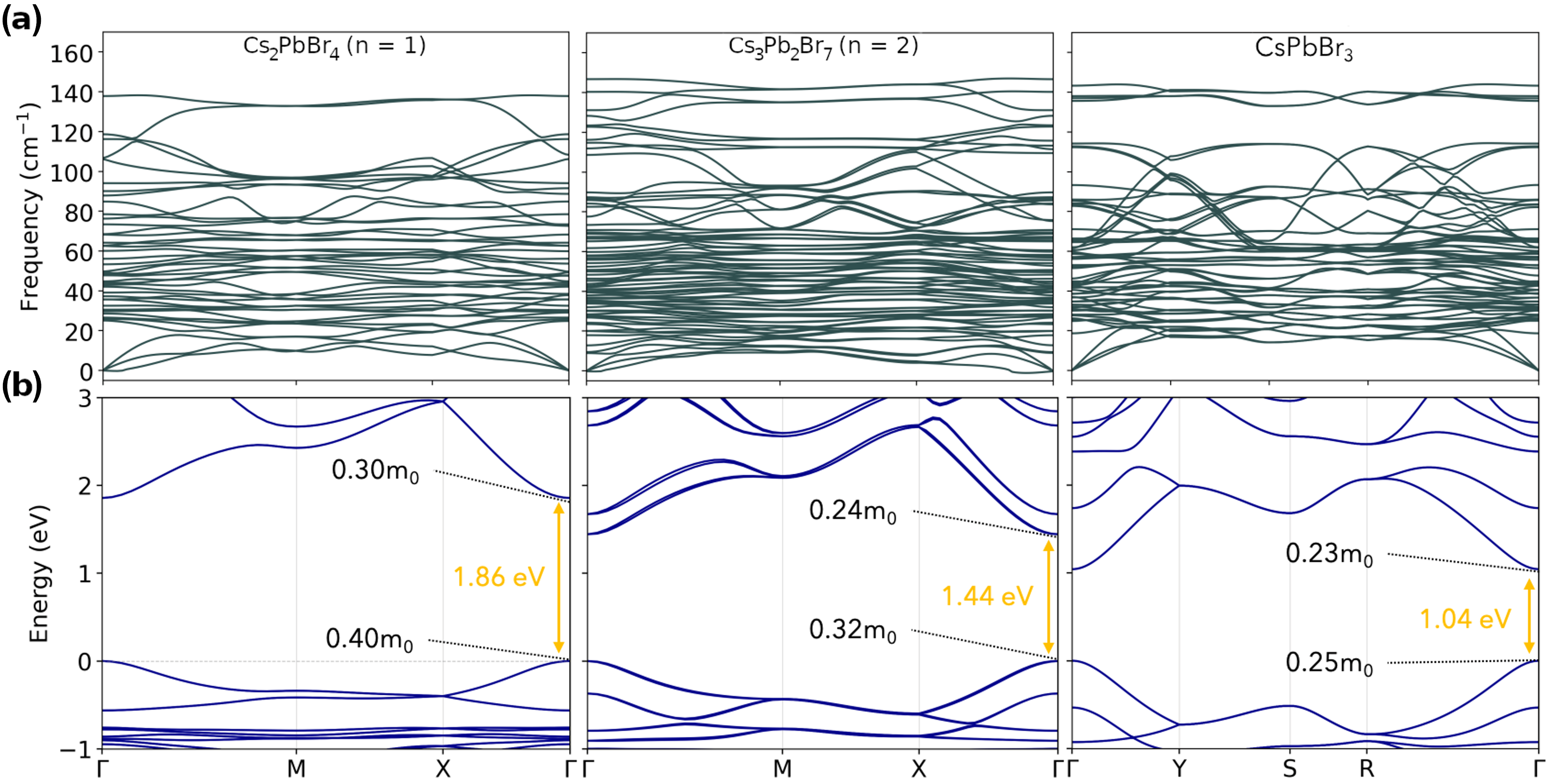}
    \vspace{-0.5cm}
    \caption{%
        Phononic and Electronic Structures.
        (a) Phonon dispersions and (b) electronic band structure for the $n=1$ and $n=2$ layered halide perovskites (Cs$_{n+1}$Pb$_{n}$Br$_{3n+1}$), and the three-dimensional CsPbBr$_{3}$. The fundamental band gaps and effective masses at band edges are provided.
        }
    \label{fig:ph-el}
\end{figure*}

We then consider the phonon properties of our models obtained by DFPT
calculations (Figure~\ref{fig:ph-el}a).
The highest phonon frequency modes around $140-150$~cm$^{-1}$ for CsPbBr$_{3}$ are in agreement with previous calculations performed on the three-dimensional cubic phase of CsPbI$_{3}$ and the orthorhombic phase of MAPbI$_{3}$, where these modes fall within the range $120-170$~cm$^{-1}$~\cite{Ponce2019}.
Importantly, the phonon dispersions confirm the mechanical stability of our models as the residual negative phonon modes are smaller than $0.01$~cm$^{-1}$.

The electronic band structure of each system is shown in Figure~\ref{fig:ph-el}b.
The orthorhombic phase of CsPbBr$_3$ exhibits a direct band gap at the $\Gamma$ point of the Brillouin zone with a computed band gap of $1.04$~eV and isotropic low hole ($0.25m_0$) and electron ($0.23m_0$) effective masses at the band edges.
When going to the layered $n=1$ model, the gap remains direct at $\Gamma$, however the band gap increases to $1.86$~eV, which constitutes a signature of the quantum and dielectric confinements.\cite{Stoumpos2016,Pedesseau2016} The top valence band and bottom conduction band remain highly dispersive within the plane of the inorganic network and the hole ($0.40m_0$) and electron ($0.30m_0$) in-plane effective masses remain low and isotropic.
With increasing number of layers, the confinement is less pronounced and the band gap for $n=2$, still direct, is decreasing to $1.44$~eV, while the in-plane effective masses becomes $0.32m_0$ and $0.24m_0$ for holes and electrons, respectively.
The similar in-plane effective masses are consistent with the similar structural parameters such as the Pb-Br bond lengths and in-plane tilting angles. 

Having obtained electronic and phonon structures, we can move to the calculation of charge carriers' mobilities. To provide insights on how the change in the effective mass between these systems affects the charge carrier mobilities, we first employ the Drude's formula, $\mu=e\tau/m^{*}$, which leads to a mobility ($\mu$) proportional to the carrier lifetime and inversely proportional to the charge carrier's effective masses ($m^*$).~\cite{Drude1900,Drude1900_2}
By assuming similar carrier lifetimes and comparing the in-plane effective masses obtained within the parabolic approximation (Figure~\ref{fig:ph-el}), it would appear that the mobilities are not drastically affected by the changes in the effective masses originated from the dimensionality reduction.
Thus, within these approximations, we conclude that by accounting only changes in the effective masses, Drude's formula cannot explain the observed mobility drop of an order of magnitude when going from three-dimensional to layered halide perovskites.

\subsection*{Ab initio Charge Carrier Mobilities}
\begin{figure}[t]
    \centering
    \includegraphics[width=0.8\linewidth]{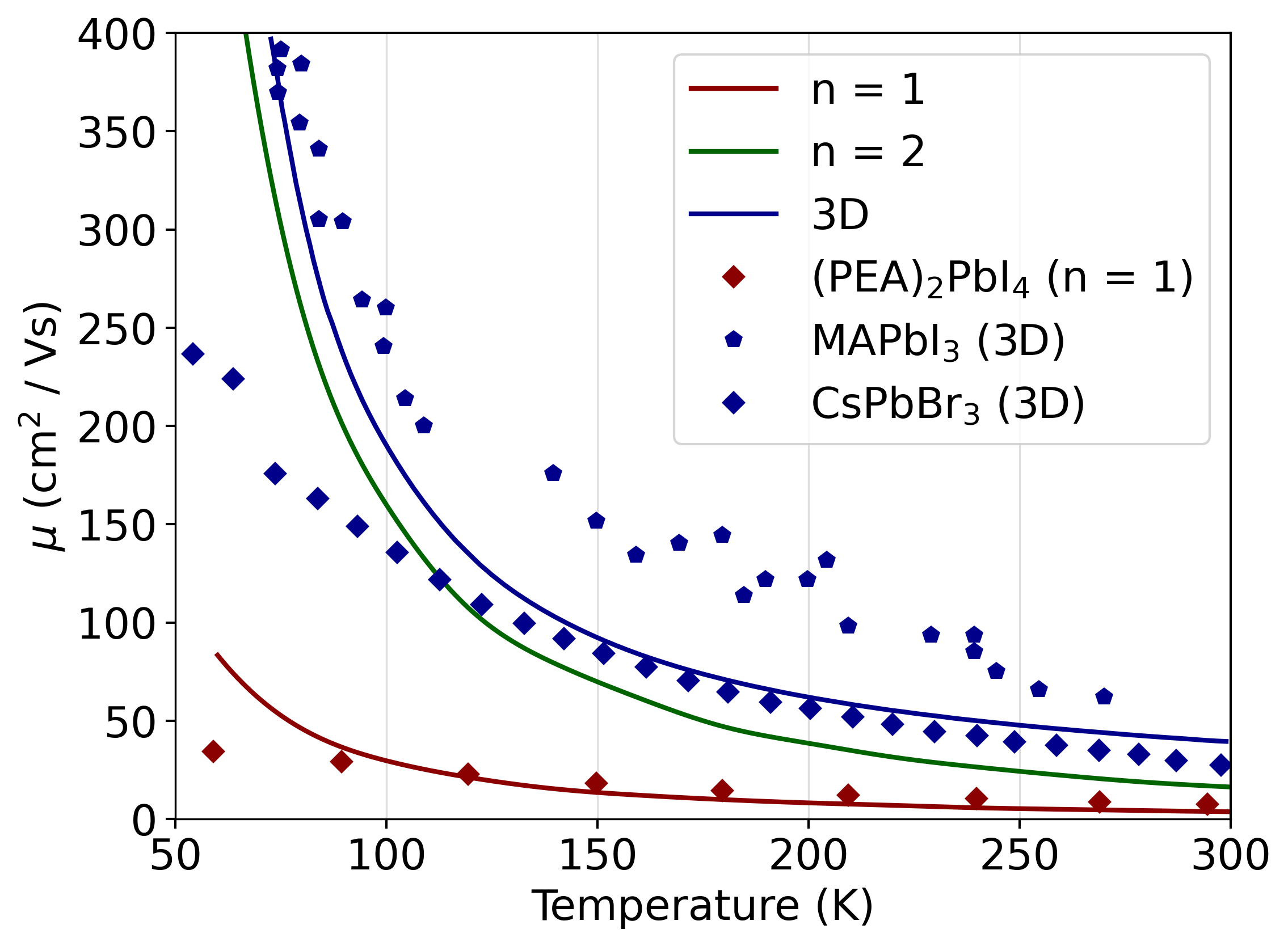}
    \vspace{-0.5cm}
    \caption{%
        Average in-plane hole mobilities.
        The calculations performed within the \textit{ai}BTE approach for the $n=1$, $n=2$, and three-dimensional systems are shown as solid lines. The experimental mobilities for the three-dimensional structures CsPbBr$_{3}$ and MAPbI$_{3}$ were extracted from references~\citenum{Bruevich2022} and~\citenum{Xia2021}, respectively. The mobility for (PEA)$_{2}$PbI$_{4}$ was extracted from Reference~\citenum{Motti2023}. The experimental values correspond to the combined electron and hole mobilities.
        }
    \label{fig:mobilities}
\end{figure}

Going beyond the Drude model, we proceed to compute charge carrier mobilities within an ab initio framework. To this aim, we express the charge carrier mobility tensor $\mu_{\alpha\beta}$ along arbitrary $\alpha$ and $\beta$ directions~\cite{Giustino2017} as follows:
\begin{equation}
    \label{eq:mobility}
    \mu_{\alpha\beta} = \frac{1}{n_{e}\Omega}\sum_{n}\int\frac{d\textbf{k}}{\Omega_{BZ}}v_{n\textbf{k},\beta}\partial_{E\alpha}f_{n\textbf{k}}.
\end{equation}
Here, \textit{n$_{e}$} is the carriers density, $\Omega$ and $\Omega_{BZ}$ are the volumes of the unit cell and Brillouin zone, respectively, $v_{n\textbf{k},\beta}$ is the velocity of an electron with crystal momentum \textbf{k} at the band \textit{n} along the $\beta$ Cartesian direction and $f_{n\textbf{k}}$ is the out-of-equilibrium carrier distribution function.
To evaluate the variation of the charge carrier distribution function in relation to the electric field along the $\alpha$ direction, $\partial_{E\alpha}f_{n\textbf{k}}$, we solve the full \textit{ai}BTE given by:
\begin{equation}
    \label{eq:carrierdist}
    \begin{aligned}
        \partial_{E\alpha}f_{n\textbf{k}} &= e\frac{\partial f_{n\textbf{k}}}{\partial \epsilon_{n\textbf{k}}}v_{n\textbf{k},\alpha}\tau_{n\textbf{k}}+\frac{2\pi}{\hbar}\tau_{n\textbf{k}}\sum_{m\nu}\int\frac{d\textbf{q}}{\Omega_{BZ}}|g_{nm\nu}(\textbf{k},\textbf{q})|^{2}\\
        &\times[(n_{\textbf{q}\nu}+1- f^{0}_{n\textbf{k}})\delta(\epsilon_{n\textbf{k}}-\epsilon_{m\textbf{k}+\textbf{q}}+\hbar\omega_{\nu\textbf{q}})\\
        &+(n_{\textbf{q}\nu}+f^{0}_{n\textbf{k}})\delta(\epsilon_{n\textbf{k}}-\epsilon_{m\textbf{k}+\textbf{q}}-\hbar\omega_{\nu\textbf{q}})]\partial_{E\alpha}f_{m\textbf{k}+\textbf{q}},
    \end{aligned}
\end{equation}
where $g_{nm\nu}(\textbf{k},\textbf{q})$ are the electron-phonon (e-ph) matrix elements associated with the scattering of an electron from the initial state $n\textbf{k}$ to the state $m\textbf{k}+\textbf{q}$ via a phonon with crystal momentum \textbf{q} in the branch $\nu$.\cite{Ponce2018} The quantities $f^{0}_{n\textbf{k}}$ and $n_{\textbf{q}\nu}$ represents the equilibrium Fermi-Dirac and Bose-Einstein distributions, respectively. Finally, the lifetimes $\tau_{n\textbf{k}}$ are obtained via the inversion of the scattering rates $\tau^{-1}_{n\textbf{k}}$. Within the Fermi golden rule it can be written as shown in equation~\ref{eq:scattering}. Phonon emission and absorption are being considered via the two terms within brackets.\cite{Lee2023,Ponce2016,Verdi2015,Giannozzi2009}
\begin{equation}
    \label{eq:scattering}
    \begin{aligned}
        \tau^{-1}_{n\textbf{k}} &= \frac{2\pi}{\hbar}\sum_{m\nu}\int\frac{d\textbf{q}}{\Omega_{BZ}}|g_{nm\nu}(\textbf{k},\textbf{q})|^{2}[(n_{\nu\textbf{q}}+1-f^{0}_{m\textbf{k}+\textbf{q}}) \\
        &\times\delta(\epsilon_{n\textbf{k}}-\epsilon_{m\textbf{k}+\textbf{q}}-\omega_{\nu\textbf{q}})\\&+(n_{\nu\textbf{q}}+f^{0}_{m\textbf{k}+\textbf{q}})\delta(\epsilon_{n\textbf{k}}-\epsilon_{m\textbf{k}+\textbf{q}}+\omega_{\nu\textbf{q}})].
    \end{aligned}
\end{equation}
The ab initio calculated in-plane hole mobilities $\mu$ are shown in Figure~\ref{fig:mobilities}. The mobilities of all materials are decreasing with the increase of temperature, due to the increase of the charge carriers scattering by lattice vibrations, i.e. electron-phonon interactions. This is by construction the sole scattering mechanism considered in our calculations. Moreover, the in-plane mobility significantly increases when going from the layered to the three-dimensional CsPbBr$_{3}$ structure, which reveals that dimensionality plays a critical role in the transport properties of perovskites. A similar phenomenon was observed by Li \textit{et al.} when studying two-dimensional (2D) and bulk InSe~\cite{Li2019} and also by Cheng \textit{et al.} for a variety of 2D semiconductors~\cite{Cheng2020}. The charge carrier mobility increases from 100~cm$^{2}$/Vs to 1000~cm$^{2}$/Vs when comparing 2D and bulk InSe. Conversely, carrier mobilities in classic III-V semiconductor quantum wells can reach very high values, thanks to quantum confinement which generate subbands and lift the degeneracy of the valence band.\cite{Blancon2020,Bastard1990} For lead bromide perovskites, at room temperature we predict hole mobilities of $4.3$~cm$^{2}$/Vs, $15.4$~cm$^{2}$/Vs and $41.6$~cm$^{2}$/Vs for the $n=1$, $n=2$ and three-dimensional CsPbBr$_{3}$ structures, respectively. The obtained mobility for CsPbBr$_{3}$ is in excellent agreement with the $41$~cm$^{2}$/Vs previously obtained by Poncé \textit{et al.} for cubic CsPbBr$_{3}$ also within the same fully ab initio approach.\cite{Ponce2019} Other sets of calculations for CsPbX$_{3}$ (X = Cl, Br, I) employing different methodologies can also be found in the literature, such as Feynman polaron models~\cite{Frost2017}, Bloch–Boltzmann theory\cite{Filippetti2016} and numerical integration of the BTE using LO electron-phonon matrix elements within the Fr{\"o}hlich model~\cite{Kang2018}. Additionally, the in-plane electron mobilities for the $n=1$ and $n=2$ layered systems are found to be $8.5$~cm$^{2}$/Vs and $16.0$~cm$^{2}$/Vs (Figure~S1 of the Support Information file (SI)~\cite{SI2024}), respectively, revealing slightly larger mobilities for electrons in agreement with Poncé \textit{et al.} results for three-dimensional halide perovskites.\cite{Ponce2019} It has been shown that the well-known DFT underestimation of the band-gap values, can lead to a slight overestimation of mobilities of up to 15\%.\cite{Ponce2021} The increase of charge carrier mobilities as a function of the number of inorganic layers is consistent with previous works by Mitzi \textit{et al.}~\cite{Mitzi1994} and Gélvez-Rueda \textit{et al.}~\cite{Rueda2017}, reporting combined electron-hole mobilities. We also note that for the $n=1$ and $n=2$ systems the mobilites were calculated using the two-dimensional form of the long-range Fr{\"o}hlich coupling as proposed by W. H. Sio and F. Giustino.~\cite{Sio2022} In Figure~S2 of the SI we show a comparison between the hole mobilities employing the two-dimensional and three-dimensional form of the long-range coupling.~\cite{SI2024} This highlights the importance of properly treating the Fr{\"o}hlich coupling for the two-dimensional compounds, due to the divergency of the el-ph matrix elements for small \textbf{q}.

\begin{figure}[t]
    \centering 
    \includegraphics[width=0.85\linewidth]{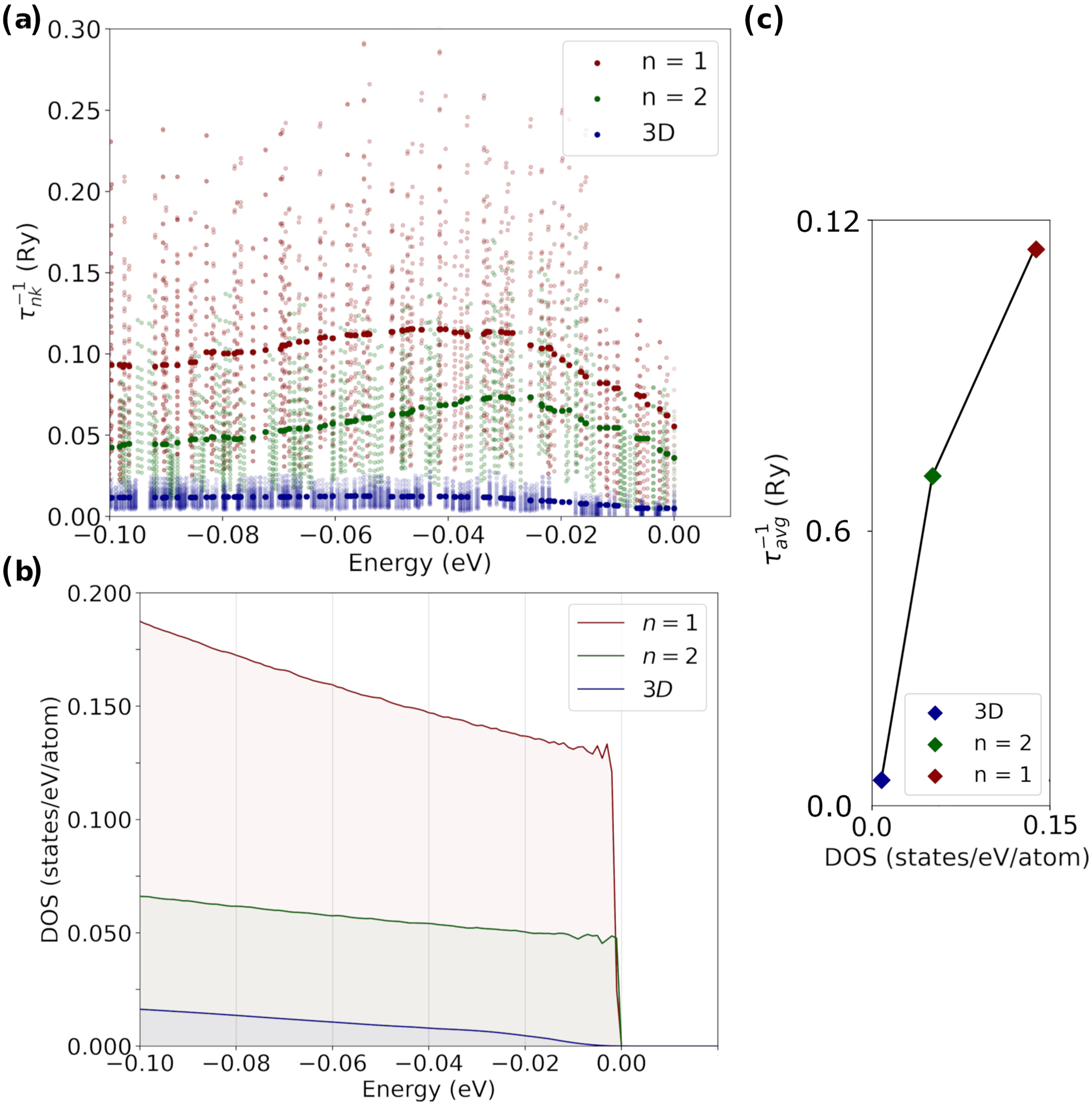}
    \vspace{-0.5cm}
    \caption{%
        Charge carrier scattering rates, electronic density of states and electron-phonon coupling strength.
        (a) Scattering rates evaluated within the \textit{ai}BTE approach. The points in dark colors represent the average scattering rates for each energy step.
        (b) Electronic density of states (DOS) near the valence band edge. The three systems' VBM are set to zero. (c) Energy-averaged scattering rate as a function of the electronic density of states. 
        }
    \label{fig:scattering}
\end{figure}

The scattering rates $\tau^{-1}_{n\textbf{k}}$ related to the hole carrier mobilities are shown in Figure~\ref{fig:scattering}a. At the valence band top the scattering rates are considerably smaller, which indicates large charge carrier lifetime. This is well-established as the carrier energies at band edges are below the threshold for LO phonon emission. The scattering rates are larger by more than an order of magnitude near the band edge for the layered $n=1$ and $n=2$ structures with respect to the three-dimensional structure, which is consistent with the hole mobilities. This indicates that the mobility trend shown in Figure~\ref{fig:mobilities} primarily originates from the different carrier lifetimes in each system.

Within time-dependent perturbation theory, the carrier's lifetimes and scattering rates can be obtained from Fermi’s golden rule.\cite{Grimvall1981} In its simplest approximation, the scattering rate is directly proportional to the electron-phonon coupling strength $|g_{nm\nu}(\textbf{k},\textbf{q})|^{2}$ and the electronic density of states D($\epsilon$) at the vicinity of the band edges, that is $\tau^{-1} \propto |g_{nm\nu}(\textbf{k},\textbf{q})|^{2}D(\epsilon)$. To unveil how the differences in the density of states of these systems contribute to the charge carrier scattering, in Figure~\ref{fig:scattering}b we show the calculated electronic density of states (DOS) near the valence band edges for the three compounds. Within this narrow energy window, the DOS is more important for the layered systems than for the three-dimensional CsPbBr$_3$. In conventional semiconductors it can also be shown that as the number of octahedra layers \textit{n} increases, the DOS of the layered systems decreases and converges to the three-dimensional limit where $D(\epsilon) \propto \sqrt{\epsilon}$.\cite{Li2019} This crossover of the electronic band structure from 2D to 3D can be understood by introducing 3D bulk-like vertical masses. However, in perovskites the electronic spacing of subbands and thus the connection between 2D and 3D DOS strongly depends on the local atomistic structure, including dangling bonds along the vertical axis.\cite{Katan2019} The increased density of states calculated for the layered materials, leads to an increased number of charge carriers being available to be scattered close to the band edges, which is the region being populated and contributing to the charge carriers' mobility at room temperature. 

In order to disentangle the contributions coming from $|g_{nm\nu}(\textbf{k},\textbf{q})|^{2}$ and the DOS, in Figure~\ref{fig:scattering}c we show the energy-averaged scattering rates (Equation~\ref{eq:tauavg}) as a function of the electronic DOS, with both quantities evaluated at the characteristic energy $3K_{b}T/2$ from the band edge: 
\begin{equation}
    \label{eq:tauavg}
    \tau^{-1}_{avg} = \frac{\sum_{nk}\tau^{-1}_{nk}\delta(\epsilon-\epsilon_{nk})}{\sum_{nk}\delta(\epsilon-\epsilon_{nk})}.
\end{equation}
We choose to evaluate Equation~\ref{eq:tauavg} at the energy $3K_{b}T/2$ as a reasonable descriptor for the carrier energy at room temperature, as previously shown by Poncé \textit{et al}.\cite{Ponce2019} We show that there is a linear relation between $\tau^{-1}_{avg}$ and $D(\epsilon)$, which demonstrate that the electron-phonon coupling strength for the three systems is similar, as $\tau^{-1} \propto |g_{nm\nu}(\textbf{k},\textbf{q})|^{2}D(\epsilon)$. These results show that the in-plane mobility of layered systems is intrinsically reduced by the more significant DOS in the vicinity of the band edges, while the electron-phonon coupling strength remains similar.

\begin{figure*}[ht]
    \centering
    \includegraphics[width=0.99\linewidth]{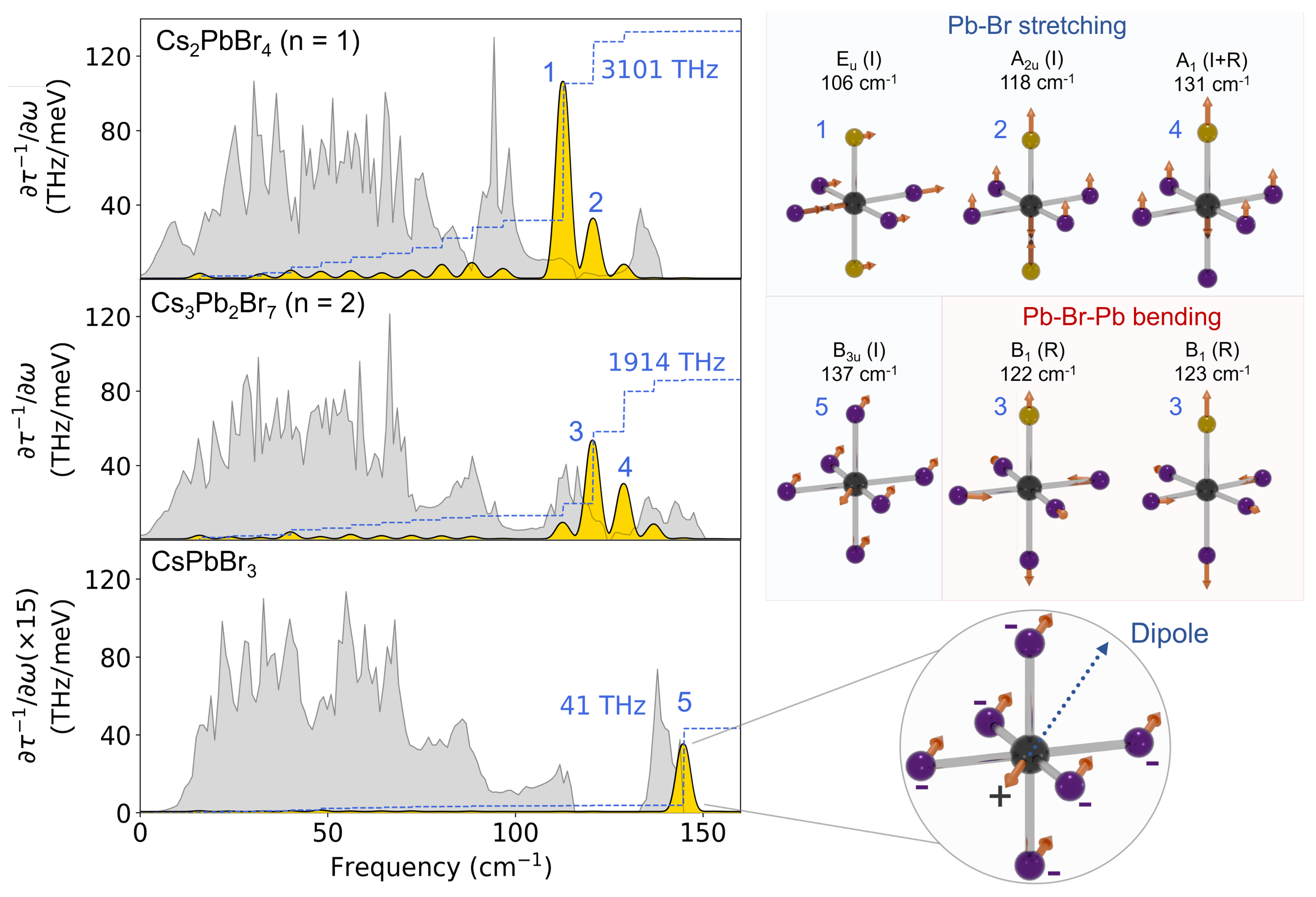}
    \vspace{-0.5cm}
    \caption{%
        Spectral decomposition of the contribution of phonons with energy $\hbar\omega$ to the scattering rates $\tau^{-1}$.
        The spectral decompositions (yellow peaks), and vibrational mode representations are show for the $n=1$ and $n=2$ layered, and three-dimensional halide perovskite. The y-axis of the three-dimensional structure plot is multiplied by fifteen for visibility. The phonon density of states is shown in gray. The cumulative integral in THz is shown in dotted blue. The atoms colored in gold represent the halogens that are not connected along the stacking direction. The gray boxes below each system indicate each system's space group and $\Gamma$-point point groups. The labels \textit{R} and \textit{I} refer to Raman-active and infrared-active, respectively.
        }
    \label{fig:spectral}
\end{figure*}

\subsection*{Spectral Decomposition of the Scattering Rates}
Having unveiled the origin of the short carrier lifetimes in the layered structures, we set to further explore the different strongly coupled phonon modes. To this aim, we analyze each phonon mode's contribution to the total scattering rate. This can be done by evaluating the variation of $\tau^{-1}$ in relation to the different phonon frequencies $\omega$, that is $\partial \tau^{-1}/\partial \omega$.
The spectral decomposition shown in Figure~\ref{fig:spectral} reveals that high-frequency optical modes dominate the scattering processes. In contrast, acoustic and low-frequency optical modes do not contribute significantly to the scattering rates. In this work we neglect the effects of the quadrupole contributions to the electron-phonon matrix elements, which could increase the coupling to ZA (Transversal out-of-plane acoustic modes), LA (Longitudinal acoustic modes) and TA (Transversal in-plane acoustic modes) phonon modes and further reduce the computed mobilities. In the case of $n=1$, the primary source of charge carrier scattering are longitudinal-optical (LO) phonons around 113~cm$^{-1}$ and 121~cm$^{-1}$ which are related to Pb-Br stretching vibrations (Peaks 1 \& 2). Due to the absence of octahedra connectivity along the stacking direction, we observe a splitting of the scattering peak associated with Pb-Br stretching modes into in-plane (Peak 1) and out-of-plane (Peak 2) contributions. Our symmetry analysis at the $\Gamma$ point for the Peak 1 mode reveals that the Pb-Br stretching mode is related to the doubly degenerated E$_{u}$ irreducible representation, which is infrared-active and is also the vectorial representation of the D$_{4h}$ point-group in the (x,y) plane. Such type of polar distortion of the lattice gives rise to an in-plane electric field which can scatter the charge carriers via the Fr{\"o}hlich interaction. Analogously, the Pb-Br stretching mode of Peak 2 is related to the A$_{2u}$ representation, which is infrared-active and is also the vectorial representation of the D$_{4h}$ point-group in the \textit{z} direction. 

For $n=2$, the scattering rate is also dominated by LO phonons associated with a similar Pb-Br stretching mode contribution (Peak 4) around 129~cm$^{-1}$ and Pb-Br-Pb bending modes both in-plane and out-of-plane (Peak 3) around 121~cm$^{-1}$. Our symmetry analysis at the $\Gamma$ point for Peak 4 shows that the stretching mode is associated with the A$_{1}$ representation which is both infrared-active and Raman-active, and is the vectorial representation of the C$_{4v}$ point-group in the \textit{z} direction. Consequently, Peak 4 observed for the $n=2$ system has the same physical origin as for $n=1$, which is related to Fr{\"o}hlich coupling. However, the symmetry analysis for the Peak 3 shows that the Pb-Br-Pb bending modes are associated with the B$_{1}$ irreducible representation, which is Raman-active, but it is not a vectorial representation of the C$_{4v}$ point-group.
Therefore, these modes can not give rise to a scattering electric field as it would be the case of a Fr{\"o}hlich interaction. We associate the scattering mechanism of Peak 3 with a non-polar optical deformation potential scattering. This type of scattering can be understood by considering the intra-band scattering limit at $q=0$ for the electron-phonon matrix elements $g_{nm\nu}(\textbf{k},\textbf{q})$, that is, $g_{nn\nu}(\textbf{k},0)=\big<u_{n\textbf{k}}|\Delta_{0\nu}v_{SCF}|u_{n\textbf{k}}\big>_{uc}$.
In this expression, if the operator $\Delta_{0\nu}v_{SCF}$ changes the symmetry of the Bloch state $|u_{n\textbf{k}}\big>$ the braket will be zero due to the orthogonality of the Bloch states, leading to zero electron-phonon coupling. The only solution leading to non-zero matrix elements are the ones where the operator preserves the symmetry of the initial Bloch state. For the $n=2$ system, by analyzing the coupling between the phonon modes with irreducible representation B$_{1}$ with the charge carriers with irreducible representation \textit{E}, we see that the transition matrices $\big<u_{n\textbf{k}}|\Delta_{0\nu}v_{SCF}|u_{n\textbf{k}}\big>_{uc}$ always contain the initial electronic state \textit{E}, which satisfies the selection rule for the deformation potential coupling.\cite{Ludwig2012} Therefore, the non-polar optical deformation potential coupling between the Pb-Br-Pb bending modes with the charge carriers is allowed by symmetry and contributes to the charge carrier scattering.

Finally, the three-dimensional system exhibit a sharp peak located at around 145~cm$^{-1}$, which also arises from the scattering via LO phonons and is related to a similar Pb-Br stretching mode as discussed for the layered systems. The symmetry analysis at the $\Gamma$ point for Peak 5 shows that the stretching mode is connected to the B$_{3u}$ irreducible representation, which is infrared-active and also the vectorial representation of the D$_{2h}$ point-group. It is important to note that the non-polar optical deformation potential scattering processes predicted for the $n=2$ system are not allowed by symmetry either for the $n=1$ or three-dimensional system. This analysis is consistent with previous symmetry analysis performed by Even \textit{et al.} for bulk cubic perovskites.\cite{Even2016} The spectral decomposition that we obtain for the bulk system is also consistent with previous calculations performed by Poncé \textit{et al.} for MAPbI$_{3}$~\cite{Ponce2019} and experimental measurements by Pérez-Osorio \textit{et al.}~\cite{Perez2015}.
This analysis demonstrate that even if the dimensionality is reduced, the primary source of charge carrier relaxation in layered perovskites is still given by polar-phonon scattering mechanisms that arise from the long-range Fr{\"o}hlich coupling, consistently with calculations for three-dimensional systems such as AMX$_{3}$ halide perovskites~\cite{Schlipf2018} or A$_{2}$MM'X$_{6}$ halide double perovskites~\cite{Leveillee2021}. However, for the layered systems, we describe additional scattering sources from phonon modes that  originate from the absence of octahedra connectivity along the stacking direction. These processes are indeed allowed by symmetry even in the maximally symmetrized reference structures of 2D perovskites.\cite{Blancon2020}

\section*{Conclusion}
In conclusion, we investigated the dimensionality crossover from three-dimensional to layered halide perovskites and provided insights into the origin of the low charge carrier mobilities in the latter. Using state-of-the-art ab initio transport calculations, we demonstrated that the decreased charge mobility observed in layered halide perovskites is primarily originating from differences in the carrier's lifetimes. By decoupling the different contributions to the charge carrier scattering rates we showed that an abrupt increase of the electronic density of states in the vicinity of the band edges for layered systems leads to increased charge scattering, and consequently poor charge transport. Conversely, the electron-phonon coupling strengh remains similar between all systems. Furthermore,
we employ electron-phonon symmetry analysis to demonstrate that as for three-dimensional perovskites, the primary source of charge carrier relaxation in layered perovskites is given by polar-phonon scattering mechanisms arising from the long-range Fr{\"o}hlich coupling. However, due to the lack of octahedra connectivity along the stacking direction for layered materials, additional vibrational modes can contribute to the scattering rates. We also show that the $n=2$ layered system can exhibit non-polar optical deformation potential scattering processes, which are not allowed by symmetry either for the three-dimensional or $n=1$ system.
Our findings provide a fundamental understanding of the electron-phonon coupling mechanisms governing the transport properties of layered metal halide perovskites and demonstrate an intrinsic limitation of the charge carrier transport in these materials. We believe that due to these limitations, favoring layered perovskite materials with larger number of \textit{n} layers would lead to higher device efficiencies in for example, 2D/3D heterostructures.
Furthermore, by developing strategies to potentially reduce the electronic DOS in the vicinity of the band edges and/or damp the identified scattering modes via for example composition engineering could potentially lead to higher performances. We are confident that this work can pave the way to the rational design of new layered perovskite-based devices and materials, exhibiting improved carrier lifetimes and more effective charge carrier separation.

\section*{Methods}
\subsection*{Density Functional Theory}
All the calculations within the DFT framework are performed using the Quantum ESPRESSO Suite.\cite{Giannozzi2017,Giannozzi2009} The layered structures are optimized with a 80~Ry cutoff for the plane-wave kinetic energy with a $4\times 4\times 1$ Brillouin zone sampling while a $4\times 4\times 4$ Brillouin zone sampling was employed for the three-dimensional structure. The threshold on forces and total energy during ionic minimization were converged to $10^{-6}$~Ry/bohr and $10^{-11}$~Ry, respectively. During the optimizations, a 0.1~Kbar threshold is employed for the pressure on the variable cell. A threshold of $10^{-11}$~Ry is employed for the SCF steps. All the calculations for hole transport properties are performed using non-relativistic norm-conserving PBE pseudopotential taken from the pseudo-dojo database (http://www.pseudo-dojo.org)~\cite{Vansetten2018}, while electron transport properties are evaluated taking spin-orbit interactions into account. The self-interaction error due to the DFT-PBE exchange-correlation functional has marginal effects on the computation of the electron-phonon matrix elements, as the charge carriers are delocalized.\cite{Giustino2017} Furthermore, although in perovskites the inclusion of spin-orbit coupling is fundamental to describe the complete valence band manifold, in the vicinity of the band edges there are no significant changes in energy and band dispersion, as shown in Figure~S3.~\cite{SI2024} The phonon calculations are performed within the DFPT framework as also implemented in Quantum ESPRESSO. A $4\times 4\times 4$ k-grid is employed for the SCF, while the phonons are solved in a $4\times 4\times 1$ q-grid for layered systems and $2\times 2\times 2$ q-grid for the three-dimensional system. A 10$^{-17}$~Ry threshold for phonon self-consistency is employed.

\subsection*{Ab Initio Boltzmann Transport Equation}
The solution of the \textit{ai}BTE was performed with the EPW code.\cite{Lee2023,Ponce2016,Verdi2015,Ponce2018,Macheda2018} For the calculation of layered systems we employ a $8\times 8\times 1$ and $4\times 4\times 1$ coarse k-grid and q-grid, respectively. These are interpolated in fine grids of $60\times 60\times 1$ for both k-grid and q-grid. The interpolation of the electronic band structure is performed using Wannier interpolation. For the Wannier interpolation of the valence and conduction band, we start with only Pb orbitals as an initial set of projectors. The resulting orbitals lead to high-quality interpolation in the narrow energy range relevant to transport. The Wannier interpolated band structure and DFT, along with the corresponding MLWFs can be found in Figure~S4 of the SI.~\cite{SI2024} We note that a more complete set would be required to interpolate the complete valence and conduction band manifold. All states within 300~meV from the band edges are included in the \textit{ai}BTE calculation. For the three-dimensional system, we employ a $4\times 4\times 4$ and $2\times 2\times 2$ coarse k-grid and q-grid, respectively. These are interpolated in fine grids of $50\times 50\times 40$ for both k-grid and q-grid. For the Wannier interpolation Pb s-orbitals and p-orbitals are employed. All states within 300~meV from the band edges are included in the \textit{ai}BTE calculation.

\section*{Data availability}
The data that supports the plots within this paper and other findings of
this study are available from the corresponding authors upon request.

\bibliography{bibliography}

\section*{acknowledgments}
 We gratefully acknowledge the discussions with Prof. V. Podzorov and Dr. V. Bruevich regarding intrinsic mobility measurements, and the fruitfull discussions with Dr. Hyungjun Lee and Prof. S. Th\'ebaud. The research leading to these results was supported by the Agence Nationale pour la Recherche through the CPJ program and the SURFIN project (ANR-23-CE09-0001). This work was granted access to the HPC resources of TGCC under the allocation 2022-A0130907682 made by GENCI. {B.C, J.L, V.H. and F.G. are supported by the Robert A. Welch Foundation under Award Number F-2139-20230405 (calculations and manuscript preparation) and the Computational Materials Sciences Program funded by the U.S. Department of Energy, Office of Science, Basic Energy Sciences, under Award No. DE-SC0020129 (EPW software development). The authors acknowledge the Texas Advanced Computing Center (TACC) at The University of Texas at Austin for providing HPC resources, including the Lonestar6 and Frontera under LRAC award 2103991, that have contributed to the research results reported within this paper. This research also used resources of the National Energy Research Scientiﬁc Computing Center, a DOE Office of Science User Facility supported by the Office of Science of the U.S.  Department of Energy under Contract No. DE-AC02-05CH11231.}

\section*{Author contributions}
B.C. data curation, formal analysis, conceptualization, investigation, methodology, writing-original draft and editing; M.K. project administration, resources, supervision, validation, writing-review and editing, conceptualization, formal analysis, methodology; G.V. project administration, resources, supervision, validation, writing-review and editing, conceptualization, formal analysis, methodology, funding acquisition; J.L. formal analysis, methodology, editing; V-A. H. formal analysis, methodology, editing; F.G. formal analysis, methodology, writing-review and editing.

\section*{Competing interests}
The authors declare no competing interests.

\section*{Additional information}
\subsection*{Supplementary information} 
The online version contains supplementary material available.

\subsection*{Correspondence} Requests for materials should be addressed to George Volonakis.
\end{document}


\title{Supplementary Information File for \\Intrinsic Limits of Charge Carrier Mobilities in Layered Halide Perovskites}

\author{Bruno Cucco}%
\affiliation{Univ Rennes, ENSCR, CNRS, ISCR (Institut des Sciences Chimiques de Rennes)– UMR 6226, F-35000 Rennes, France.}%
\affiliation{Oden Institute for Computational Engineering and Sciences, The University of Texas at Austin, Austin, Texas 78712, USA
and Department of Physics, The University of Texas at Austin, Austin, Texas 78712, USA.}%
\author{Joshua Leveillee}%
\affiliation{Oden Institute for Computational Engineering and Sciences, The University of Texas at Austin, Austin, Texas 78712, USA
and Department of Physics, The University of Texas at Austin, Austin, Texas 78712, USA.}%
\author{Viet-Anh Ha}%
\affiliation{Oden Institute for Computational Engineering and Sciences, The University of Texas at Austin, Austin, Texas 78712, USA
and Department of Physics, The University of Texas at Austin, Austin, Texas 78712, USA.}%
\author{Jacky Even}%
\affiliation{Univ Rennes, INSA Rennes, CNRS, Institut FOTON - UMR 6082, F-35000 Rennes, France.}%
\author{Mika\"el Kepenekian}%
\affiliation{Univ Rennes, ENSCR, CNRS, ISCR (Institut des Sciences Chimiques de Rennes)– UMR 6226, F-35000 Rennes, France.}%
\email{mikael.kepenekian@univ-rennes.fr}
\author{Feliciano Giustino}%
\affiliation{Oden Institute for Computational Engineering and Sciences, The University of Texas at Austin, Austin, Texas 78712, USA
and Department of Physics, The University of Texas at Austin, Austin, Texas 78712, USA.}%
\author{George Volonakis}%
\email{yorgos.volonakis@univ-rennes.fr}
\affiliation{Univ Rennes, ENSCR, CNRS, ISCR (Institut des Sciences Chimiques de Rennes)– UMR 6226, F-35000 Rennes, France.}%

\maketitle

\newpage
\clearpage
\setcounter{figure}{0}
\setcounter{table}{0}
\makeatletter 
\renewcommand{\thefigure}{S\@arabic\c@figure}
\makeatother
\pagenumbering{arabic}
\makeatletter 
\renewcommand{\thetable}{S\@arabic\c@table}
\makeatother

\clearpage
\newpage

\begin{figure}[]
 \begin{center}
  \includegraphics[width=0.75\linewidth]{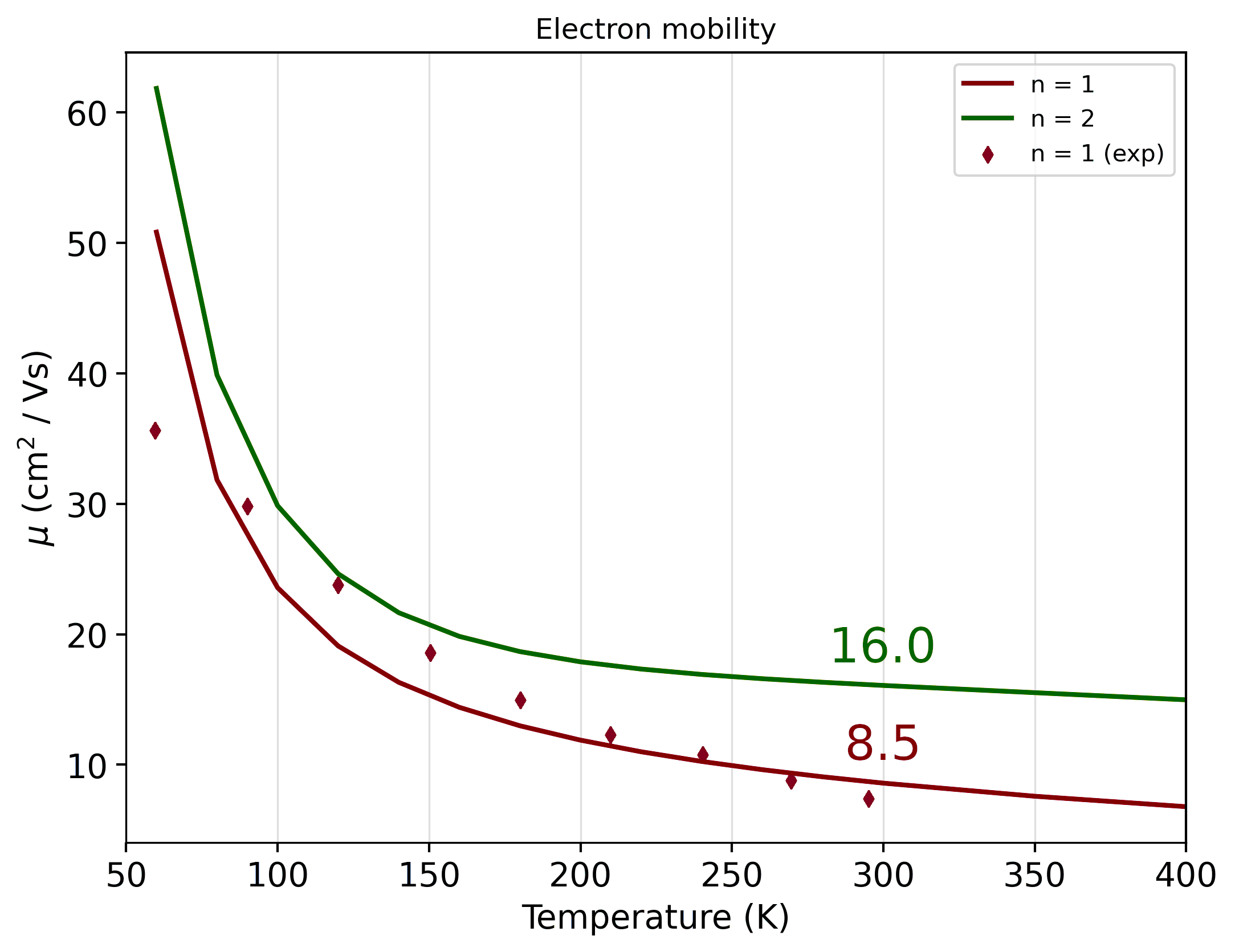}
 \end{center}
 \vspace{-0.5cm}
 \caption{\textbf{Average in-plane electron mobilities.} Calculations performed within the \textit{ai}BTE approach for the $n=1$ and $n=2$. The experimental mobilities for the layered $n=1$ were extract from reference~\cite{Motti2023}.}
 \label{fig:mobelec}
\end{figure}

\clearpage
\newpage
\begin{figure}[]\newpage
 \begin{center}
  \includegraphics[width=0.65\linewidth]{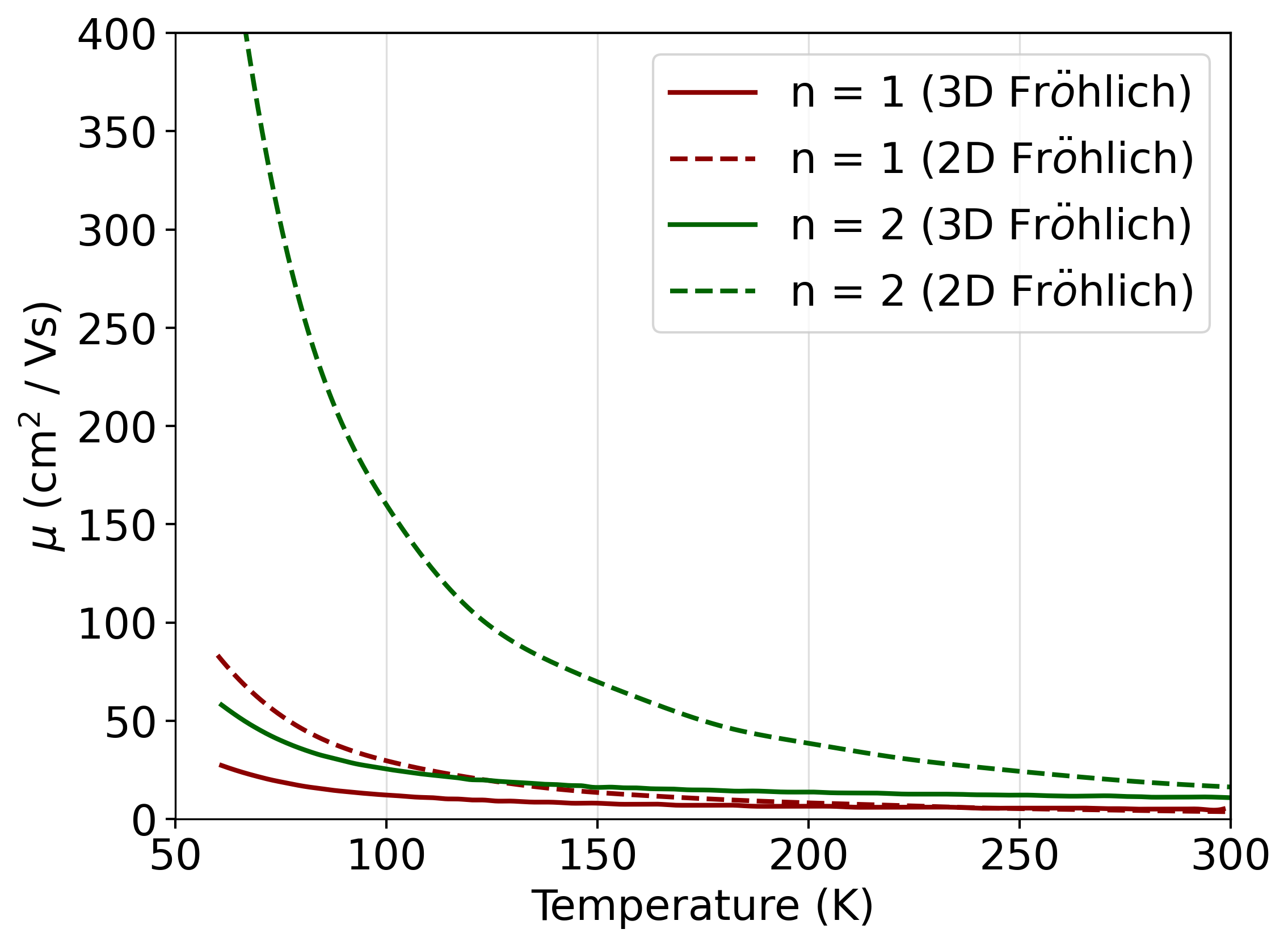}
 \end{center}
 \vspace{-0.5cm}
 \caption{\textbf{Average in-plane hole mobilities using the 3D and 2D form of the Fr{\"o}hlich coupling.} Calculations performed within the \textit{ai}BTE approach for the $n=1$ and  $n=2$ system using the two-dimensional and three-dimensional functional form of the Fr{\"o}hlich coupling.\cite{Sio2022} }
 \label{fig:2dfroh}
\end{figure}

\noindent
In Fig. S2 we show that employing the two-dimensional form of the Fr{\"o}hlich coupling leads to signicantly higher mobilities at low temperature, while at room temperature we observe negligible effects. This is attributed to the divergency of the el-ph matrix elements as $1/|\textbf{q}|$ for small \textbf{q} when the three-dimensional form is employed, which leads to larger scattering rates accordingly to Equation~4 of the main text. Consequently, lower mobilities are calculated, while for the two-dimensonal form the matrix elements are finite at small \textbf{q} leading to larger mobilities.

\clearpage
\newpage
\begin{figure}[]\newpage
 \begin{center}
  \includegraphics[width=0.55\linewidth]{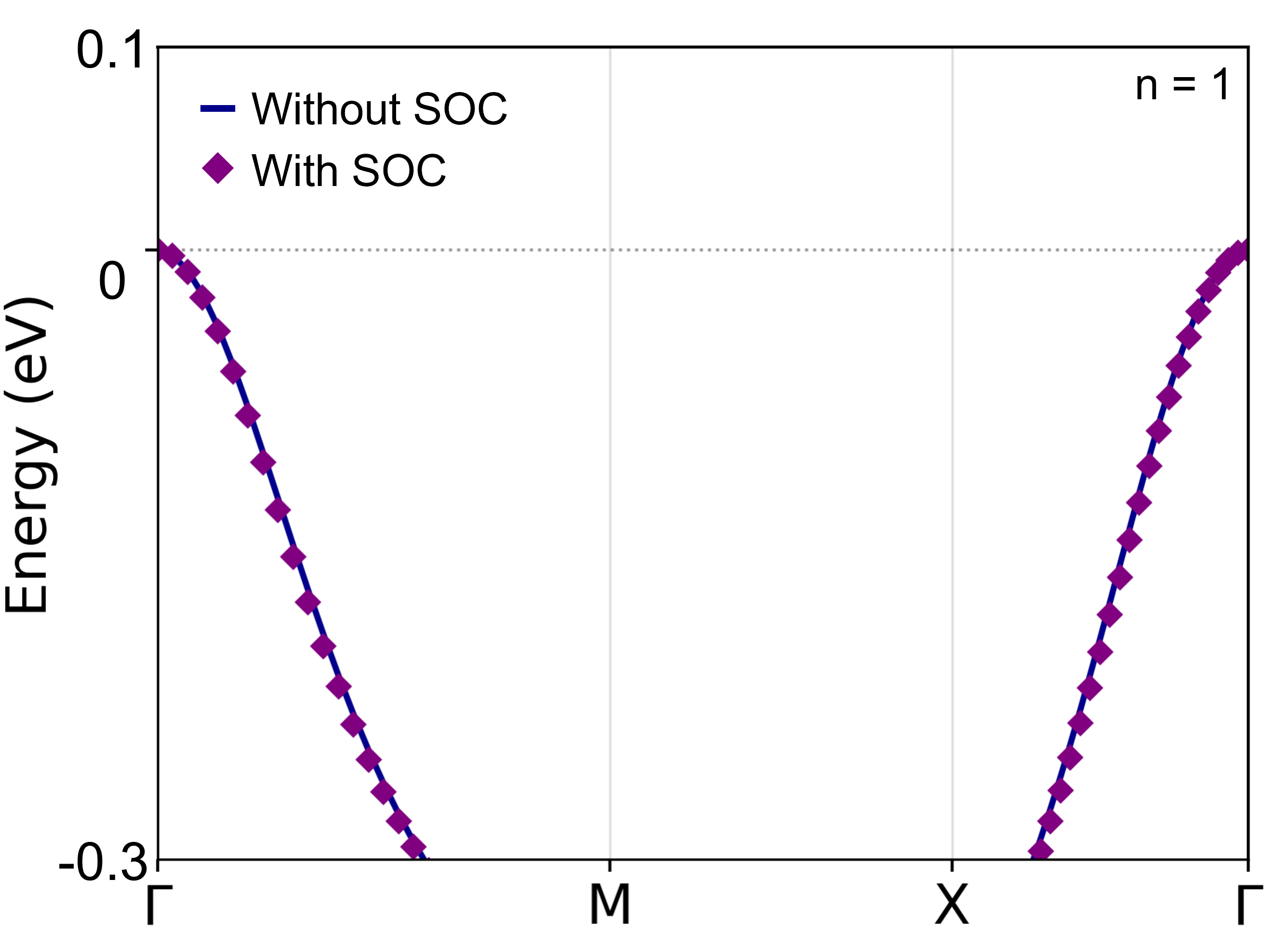}
 \end{center}
 \vspace{-0.5cm}
 \caption{\textbf{Valence band structure of $n=1$ close to the band edges.} Comparison between the valence band edge of $n=1$ with and without spin-orbit coupling.}
 \label{fig:soc}
\end{figure}

\clearpage
\newpage
\begin{figure}[]\newpage
 \begin{center}
  \includegraphics[width=1\linewidth]{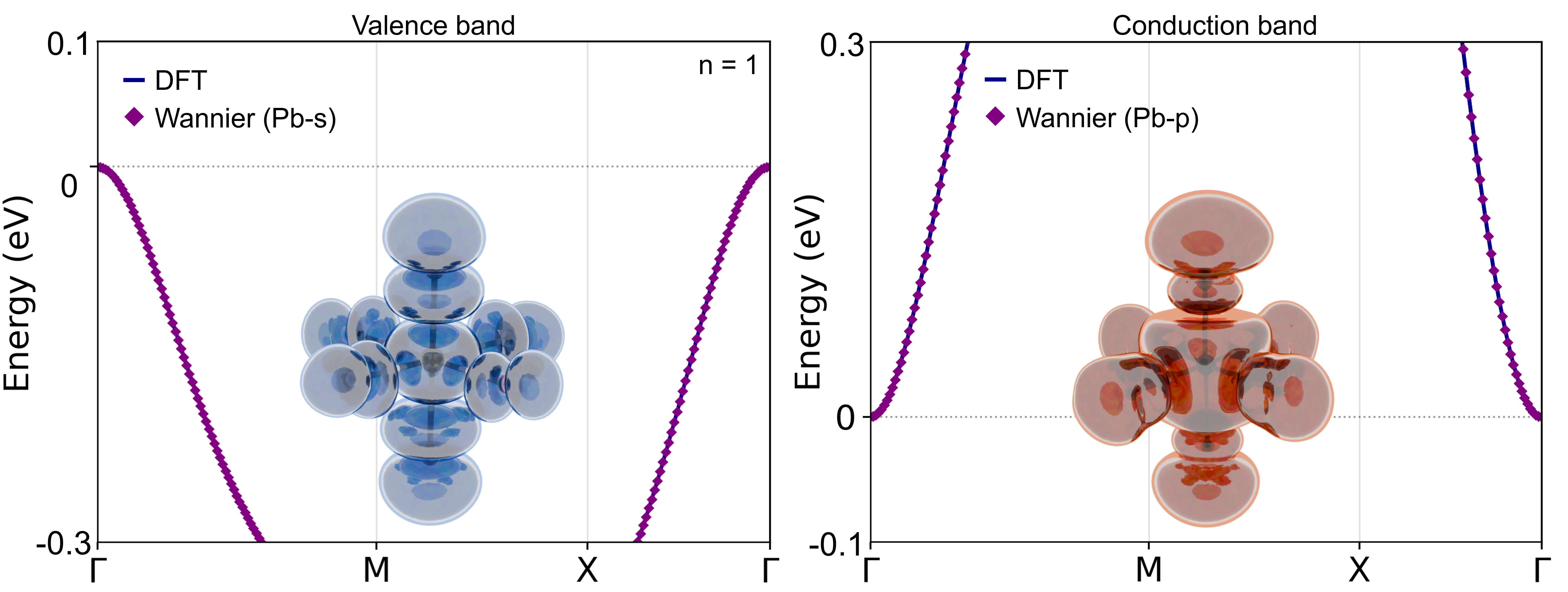}
 \end{center}
 \vspace{-0.5cm}
 \caption{\textbf{Interpolated band structure of $n=1$ close to the band edges.} Comparison between the Wannier interpolation of valence and conduction band edges concerning the DFT band structure. The real space representation of the MLWF is also shown.}
 \label{fig:wannier}
\end{figure}

\newpage
\nocite{*}
\bibliography{sibiblio}